\documentclass[epj]{webofc}
\usepackage[utf8]{inputenc}
\usepackage[varg]{txfonts}   
\usepackage{booktabs}
\usepackage{xcolor}
\definecolor{darkred}{rgb}{0.4,0.0,0.0}
\definecolor{darkgreen}{rgb}{0.0,0.4,0.0}
\definecolor{darkblue}{rgb}{0.0,0.0,0.4}
\usepackage[bookmarks,linktocpage,colorlinks,
    linkcolor = darkred,
    urlcolor  = darkblue,
    citecolor = darkgreen]{hyperref}
\usepackage{subfigure}
\wocname{EPJ Web of Conferences}
\woctitle{Lattice2017}
\begin{document}
%
\selectlanguage{english}
\title{%
Charmonium-nucleon interactions from the time-dependent HAL QCD method
}
\author{%
\firstname{Takuya} \lastname{Sugiura}\inst{1}\fnsep\thanks{Speaker, \email{sugiura@rcnp.osaka-u.ac.jp}} \and
\firstname{Yoichi} \lastname{Ikeda}\inst{1} \and
\firstname{Noriyoshi} \lastname{Ishii}\inst{1}
}
\institute{%
Research Center for Nuclear Physics (RCNP), Osaka University, Osaka 567-0047, Japan
}
\abstract{%
The charmonium-nucleon effective central interactions have been
computed by the time-dependent HAL QCD method.  This gives an updated
result of a previous study based on the time-independent method, which
is now known to be problematic because of the difficulty in achieving
the ground-state saturation.  We discuss that the result is consistent
with the heavy quark symmetry. No bound state is observed from the
analysis of the scattering phase shift; however, this shall lead to a
future search of the hidden-charm pentaquarks by considering
channel-coupling effects.
}
\maketitle
\section{Introduction}\label{sec:introduction}

Properties of hadrons having more complex structure than those of the
standard three-quark baryons or the quark-antiquark mesons, called
exotic hadrons, have not been established yet.
Recently, the LHCb collaboration have reported the observation of two
hidden-charm pentaquark candidates, $P_c(4380)$ and $P_c(4450)$, in
the $J/\psi p$ invariant mass distribution from the weak decay of the
$\Lambda_b^0 \to J/\psi p K^-$ \cite{Aaij:2015tga}.
The fit with the Breit-Wigner parameterization shows the masses and
widths as $M_{P_c(4380)}=205\pm18\pm86 \mathrm{MeV}$,
$\Gamma_{P_c(4380)}=205\pm18\pm86 \mathrm{MeV}$,
$M_{P_c(4450)}=4449.8\pm1.7\pm2.5 \mathrm{MeV}$, and
$\Gamma_{P_c(4450)}=39\pm5\pm19 \mathrm{MeV}$.  Three acceptable
spin-parity combinations $(3/2^-,5/2^+), (3/2^+,5/2^-), (5/2^+,
3/2^-)$ are also obtained.

\begin{figure}[thb]
\begin{minipage}{0.5\hsize}
  \centering
  \includegraphics[width=5cm]{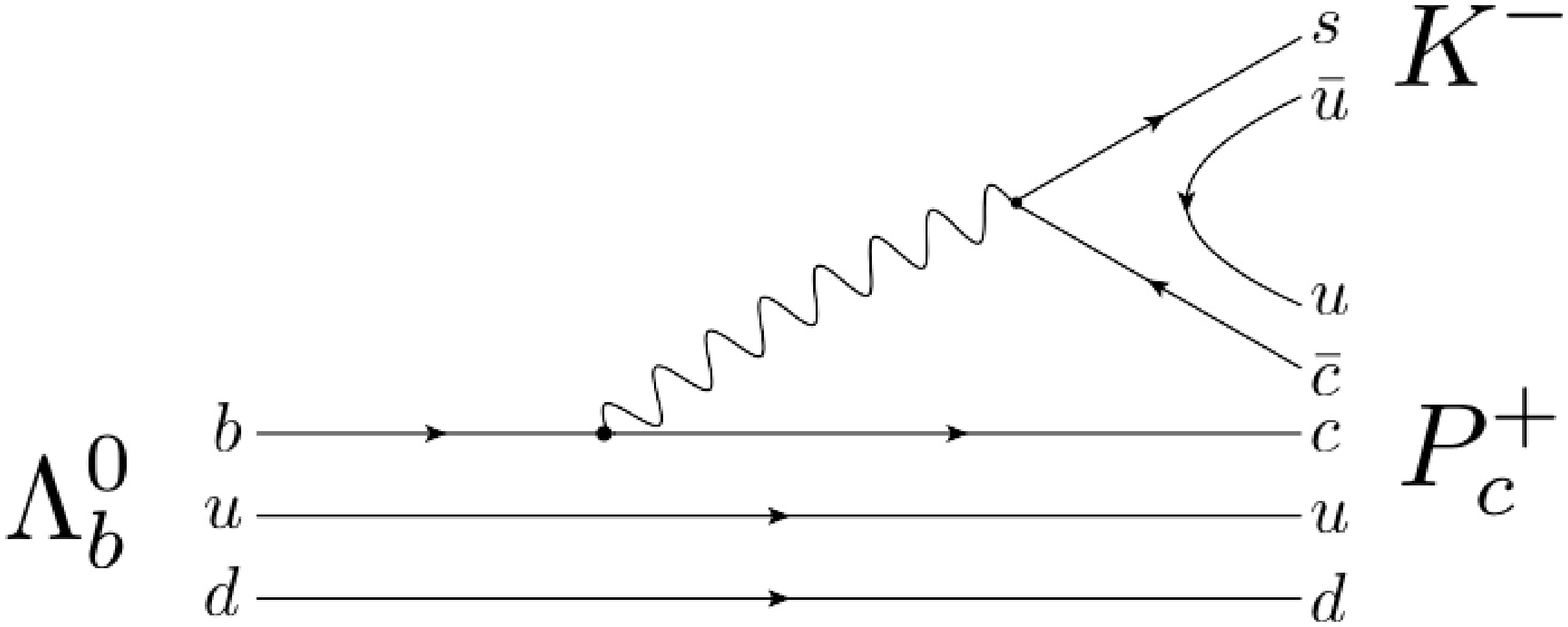}
  \label{fig:diagram}
  \caption{
    Feynman diagram for the $\Lambda_b^0 \to P_c^+ K^-$ decay.
  }
\end{minipage}
\begin{minipage}{0.5\hsize}
  \centering
  \includegraphics[width=5cm]{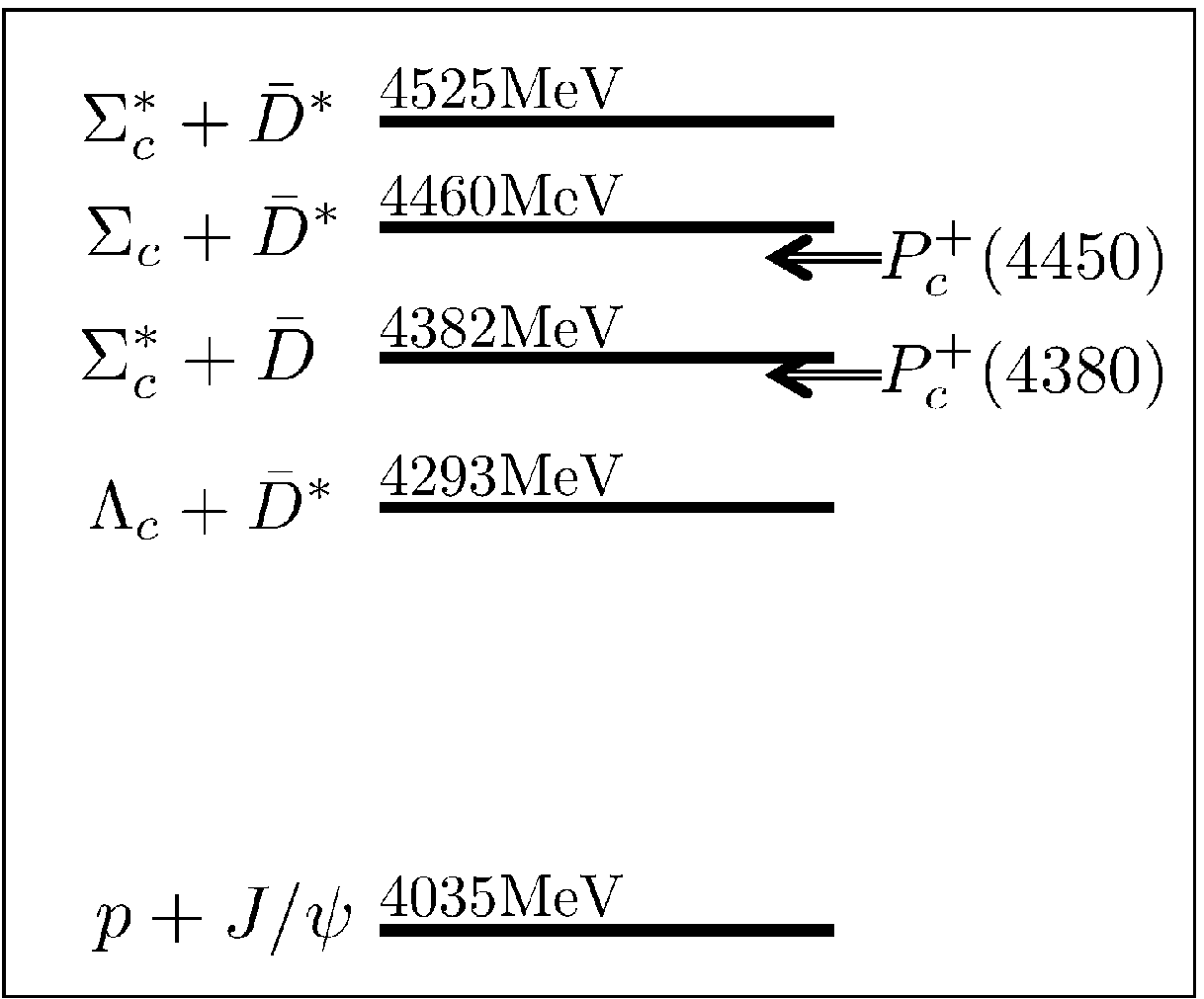}
  \label{fig:coupled_channels}
  \caption{
    The five S-wave meson-baryon channels that possibly couple to the
    pentaquarks with $J^P=3/2^-$.
  }
\end{minipage}
\end{figure}

A lot of theoretical works have been done to reveal the structure of
the
$P_c$'s~\cite{Wu:2010jy,Chen:2015loa,Chen:2015moa,Maiani:2015vwa,Guo:2015umn}.
These results do not give a consistent answer about what the reported
$P_c$ resonances actually are like, and even the existence of such
resonances.  We believe this situation originates from the lack of
knowledge of interactions between charmed hadrons.
Shown in figure~\ref{fig:diagram} is a diagram for the decay
$\Lambda_b \to P_c K$. The observed peaks of $P_c$ must couple
to the $J/\psi N$ were they really resonances. 
Therefore, it would be desirable to consider the 2-body scattering
without the spectator $K$ to study $P_c$.  Obviously such an
experiment is unaccessible, and another approach is in demand.

To settle the above issue, we aim at studying the structures of the
$P_c$'s via the lattice QCD simulations without requiring further
experimental data. We employ the potential approach developed by the
HAL QCD collaboration\cite{Ishii:2006ec}.
The method, which we shall call {\it the (time-independent) HAL QCD
  method}, utilizes the spatial part of hadron correlators, called the
Nambu-Bethe-Salpeter (NBS) amplitude.  The long-distance asymptotic
behavior of the NBS amplitude is identical to that of a scattered wave
function in quantum mechanics; therefore, by solving the Schr\"odinger
equation for the potential with such an amplitude as input, we can be
sure that the potential is faithful to the QCD S matrix.
It has already been applied to a variety of systems, including the
nucleon-nucleon (NN), nucleon-hyperon (NY), YY, and NNN interactions
\cite{Aoki:2012tk}.
The above method relies on the ground-state saturation in correlation
functions. However, the ground-state saturation has turned out
difficult to achieve in actual calculations. To avoid this problem,
the HAL QCD collaboration has introduced another method, called {\it
  the time-dependent method} \cite{HALQCD:2012aa}.
In the time-dependent method, we assume that correlation functions are
dominated not only by the ground state but by the elastic states.  By
solving a time-dependent Schr\"odinger-like equation, we calculate the
same potential as in the original method.  Since the elastic-state
saturation is much easier to achieve than the ground-state saturation,
the time-dependent method should always be preferred.
Moreover, the HAL QCD method is straightforwardly applied to
coupled-channel systems \cite{Aoki:2012tk}.  Since channel-coupling is
expected to be crucial to study the origin of the peak corresponding
to the $P_c$'s, this feature is quite helpful.

In the present study, we have set out on our coupled-channel analysis
from the $J/\psi N$ and $\eta_c N$ single-channel interactions.  We
focus on the S-wave angular momentum, so we have three channels:
$J/\psi N$ ($J^P=1/2^-$), $J/\psi N$ ($J^P=3/2^-$), and $\eta_c N$
($J^P=1/2^-$).
These channels are a part of the coupled channels that can couple to
$P_c$. The $J/\psi N$ ($J^P=3/2^-$) in particular, is the lowest-lying
channel of the five S-wave meson-baryon channels in
Figure~\ref{fig:coupled_channels}.
We calculate the effective central potential for each of the three
channels, so that the couplings to the other meson-baryon channels, to
the channels with higher angular momentum, and between $J/\psi N$ and
$\eta_c N$ in the $J=1/2$ case, are all integrated out.
The charmonium-nucleon effective central interactions have already
been studied with the quenched lattice QCD in
Ref.~\cite{Kawanai:2010ev}.
They found no bound state between $\eta_c N$ nor $J/\psi N$.
In contrast, authors in Ref.~\cite{Beane:2014sda} have reported
contradictory results from the direct method: the $\eta_c N$ system
has a deeply bound state with binding energy of $19.8\mathrm{MeV}$.
In this way, there exists a conflict between the HAL QCD potential
method and the direct method as in the NN sector where the
ground state saturation plays a key role \cite{Iritani:2017rlk}.
Since the authors of Ref.~\cite{Kawanai:2010ev} have employed the
old-fashioned HAL QCD method ({\it i.e.} the time-independent method)
which can be afflicted with the problem of the ground state
saturation,
we shall improve their results by employing the new one ({\it i.e.}
the time-dependent HAL QCD method) to avoid the problem.
Moreover, we shall see that the comparison of our results for the two
spin assignments of the $J/\psi N$ gives an implication of the heavy
quark spin symmetry.

\section{Method}\label{sec:method}

We start from computing a normalized four-point correlation function
$R$ of each charmonium-nucleon system,
\begin{align}
\label{eq:R-correlator}
  R(\vec{r},t) 
  \equiv 
  \langle 0 | T N(\vec{x}+\vec{r},t) \phi(\vec{x},t) \overline{\mathcal{J}}(t_0=0) | 0 \rangle 
  / e^{-(m_N+m_\phi)t},
\end{align}
where $N(\vec{x},t)$ denotes an interpolating operator for the
nucleon, $\phi(\vec{x},t)$ for the charmonium (either $J/\psi$ or
$\eta_c$), and $\overline{\mathcal{J}}(t_0)$ the corresponding
wall-source operator.  In actual lattice calculations, the
exponential factor involving the nucleon mass $m_N$ and the charmonium
mass $m_\phi$ is replaced by the single-hadron correlators.
The non-local potential $U(\vec{r},\vec{r}^\prime)$ satisfies the
Schr\"odinger-like equation at $t$ large enough for the elastic-state
saturation
\begin{align}
\label{eq:shrodinger}
\left( 
    - \frac{\partial}{\partial t} 
    + \frac{\vec{\nabla}^2}{2\mu}
\right)R(\vec{r},t)
=
\int d^3 \vec{r}^\prime \,
U(\vec{r},\vec{r}^\prime) R(\vec{r}^\prime, t),
\end{align}
where $\mu=1/(1/m_N+1/m_\phi)$ is the reduced mass.  We neglect the
relativistic effects, which in principle appear as higher-order time
derivatives on the left-hand side, since they are much smaller than
the other terms.
The non-locality of the potential $U(\vec{r},\vec{r}^\prime)$ can be
well-tamed by the derivative expansion, such that the scattering phase
shift computed from the expanded potential approaches the exact
one order by order~\cite{Sugiura:2017vwo}.
By taking the leading order term of the derivative expansion,
$U(\vec{r},\vec{r}^\prime)\simeq V(r)\delta^3(\vec{r}-\vec{r}^\prime)$,
we get the effective central potential
\begin{align}
\label{eq:each_term}
V_{\text{eff}}(r) = 
  - \frac{\partial_t R(\vec{r},t)}{R(\vec{r},t)}
  + \frac{1}{2\mu} \frac{\vec{\nabla}^2 R(\vec{r},t)}{R(\vec{r},t)}
\equiv V_{t}(r) + V_{\nabla}(r).
\end{align}
Note that $V_{\text{eff}}(r)$ does not depend on $t$ although the $R$
correlator does. This is true when the elastic-state saturation is
achieved and the local potential well approximates the non-local
kernel $U(\vec{r},\vec{r}^\prime)$.  Furthermore, when the
ground-state saturation is achieved, $V_t(r)$ is reduced to just a
constant, $V_t(r)=\epsilon$ with $\epsilon$ independent of both $t$
and $r$.  In other words, when $V_t(r)$ is dependent on $r$, the
time-independent method will presumably fail.

\section{Simulation Setup}\label{sec:setup}

We employ the $2+1$ flavor full QCD gauge configurations by the
CP-PACS and JLQCD Collaborations \cite{cppacs_jlqcd}.  They are
generated with the renormalization-group improved gauge action and a
non-perturbatively $\mathcal{O}(a)$ improved clover quark action at
$\beta=1.83$ (corresponding to the lattice spacing of
$a=0.1209\mathrm{fm}$ \cite{Ishikawa:2007nn}) on a $16^3\times 32$
lattice, so that the spatial volume is $(1.93\mathrm{fm})^3$.
The heavy mass of charm quarks may bring large discretization errors,
for which the relativistic heavy quark (RHQ) action~\cite{Aoki:2001ra}
should be employed. However, in this present study, we use the clover
action for all quarks.
The hopping parameter for the light (up and down) quarks and the
strange quark are set to $\kappa_{ud}=0.13760$ and $\kappa_s=0.13710$,
respectively.
For the charm quark, the value $\kappa_c=0.11660$ is determined by
fitting the $1/\kappa$ vs $m_{\text{PS}}^2$ plot to reproduce the
physical $\eta_c$ mass of $2983\mathrm{MeV}$ on this lattice.

The Dirichlet boundary condition is imposed at $t-t_0=26$ to prevent
inverse propagation. In the spatial directions we imposed the periodic
boundary condition.
We average the results from $16$ source positions $t_0$ to reduce
statistical error.
The spatial correlation is averaged based on the $48$ octahedral group
transformations.

\section{Results and Discussion}\label{sec:Results}

\begin{figure}[thb] 
\begin{minipage}{0.5\hsize}
  \centering
  \includegraphics[width=6.5cm]{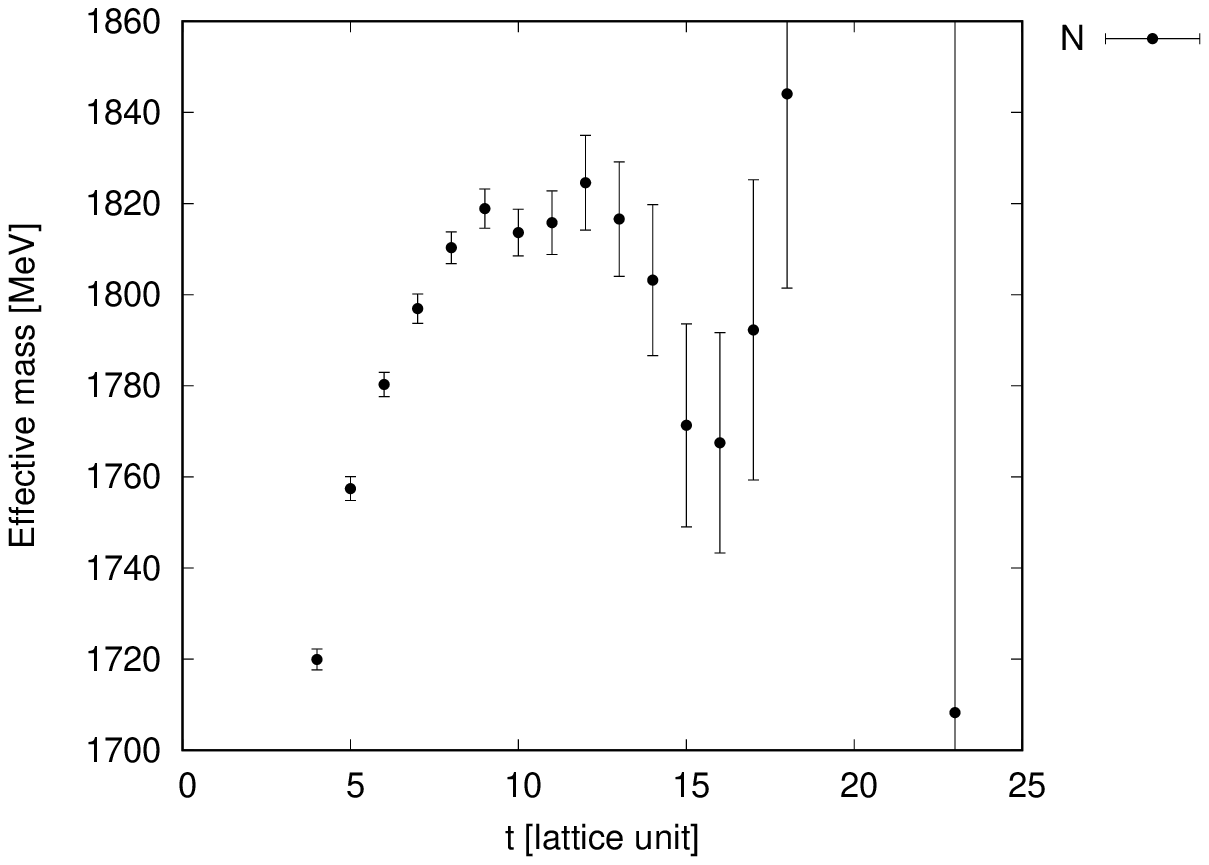}
\end{minipage}
\begin{minipage}{0.5\hsize}
  \centering
  \includegraphics[width=6.5cm]{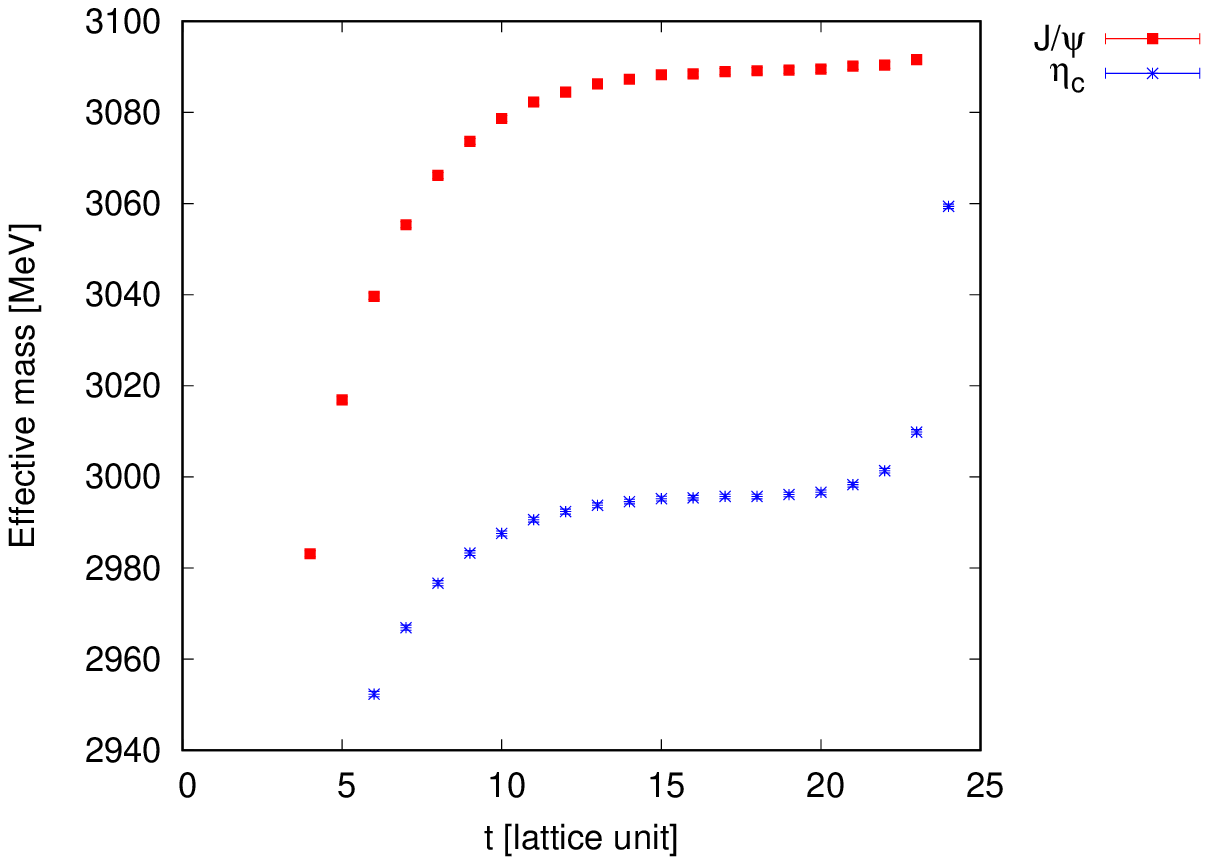}
\end{minipage}
  \caption{
    \label{fig:effective_mass}
    The effective masses of (left) nucleon, (right,blue) $\eta_c$,
    and (right,red) $J/\psi$.
  }
\end{figure}

In Figure~\ref{fig:effective_mass}, we show the effective masses of
$N$, $\eta_c$, and $J/\psi$ extracted from single-hadron correlation
functions.  We see plateaux in all those plots.
We have performed a single exponential fit to each of the
single-hadron correlators. The resulting masses are
$m_N=1816\mathrm{MeV}$ ($t=8-14$), $m_{\eta_c}=2996\mathrm{MeV}$
($t=15-18$), and $m_{J/\psi}=3089\mathrm{MeV}$ ($t=16-19$), where the
time slices used for the fit are shown in the parentheses.

\begin{figure}[thb]
  \centering
  \includegraphics[width=8cm]{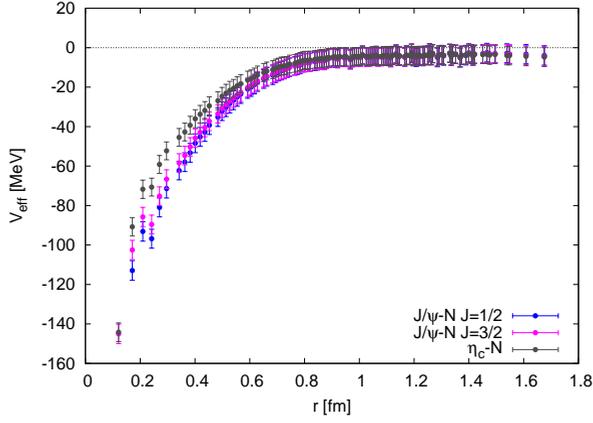}
  \caption{
    \label{fig:pot_all}
    Effective central potentials at $t=12$ of (black)$\eta_c N$,
    (blue)$J/\psi N$ with $J=1/2$, and (magenta)$J/\psi N$ with
    $J=3/2$.
}
\end{figure}

Figure~\ref{fig:pot_all} shows our results of the effective central
potentials at $t=12$.  They are all attractive overall, and are quite
similar quantitatively.
Especially, the two total spin assignments of $J/\psi N$ have very
small difference, which differ in the spin direction of $J/\psi$
relative to that of $N$, when the S-wave angular momentum considered.
This may be because of the following reason.
According to the heavy quark spin symmetry~\cite{Isgur:1991xa}, light
quarks and soft gluons see a heavy quark as a static color source,
since they cannot change the velocity or flip the spin of the heavy
quark. Thus the dynamics is independent of the heavy quark spin, as
the difference is suppressed by $\mathcal{O}(1/m_h)$ for heavy quark
with mass $m_h$.  This picture seems to be applicable to the two spin
assignments of the $J/\psi N$.
The $\eta_c N$ potential has a little weaker attraction than the
$J/\psi N$ potentials. This might be because of the lighter mass of
$\eta_c$ than of $J/\psi$ by $93\mathrm{MeV}$.
The authors of Ref.~\cite{Kawanai:2010ev} have also observed a similar
tendency between the $\eta_c N$ and spin-averaged $J/\psi N$
potentials.  They have discussed that the difference in the reduced
masses is too small to account for the $\mathcal{O}(10)\%$ difference
in the potentials, and thus it is caused by different structures of
$\eta_c$ and $J/\psi$.
We cannot draw a definite conclusion either; however we remark that
the similarity in the behavior of the charmonium-nucleon interactions,
and yet the small difference between the $\eta_c N$ and $J/\psi N$,
are compatible with the heavy quark symmetry.

\begin{figure}[thb]
  \centering
  \includegraphics[width=8cm]{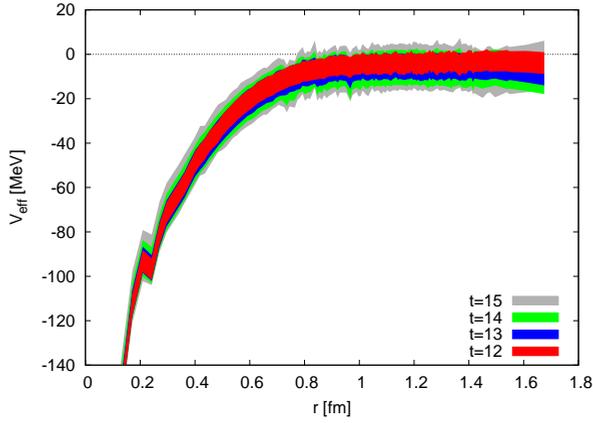}
  \caption{
    \label{fig:pot_tdep}
    The $J/\psi N$ ($J=1/2$) potentials extracted at different time
    slices, $t=12-15$.
  }
\end{figure}

In figure~\ref{fig:pot_tdep}, we show the $J/\psi N$ ($J=1/2$)
potentials extracted at $t=12-15$, where the filled areas express
statistical error bars.
We see all the results agree within the errors.  Therefore we conclude
that the elastic-state saturation is achieved and the derivative
expansion is converged at the lowest order at $t=12$.
The same check has been done to the other two potentials, and they are
found qualitatively similar to this case.

Notice that the $R$ correlator in Eq.~\eqref{eq:R-correlator} is
normalized by a single exponential factor, so that we assume the
ground-state saturation of the $J/\psi$ and $N$ single-hadron
correlators.
As can be seen in figure~\ref{fig:effective_mass}, however, $t=12$ is
a little too small for the single-charmonium (both $\eta_c$ and
$J/\psi$) correlators to achieve the ground-state saturation.
It results in $\sim 3\mathrm{MeV}$ deviation in the effective masses
of the charmonia at $t=12$ and those extracted from the plateaux in
$t>15$.
Nevertheless, the deviation is of $\sim0.1\%$ relative to the total
masses of $\sim 3000\mathrm{MeV}$, so that it affects little to the
potentials, as is actually seen in Figure~\ref{fig:pot_tdep}.

\begin{figure}[thb]
  \centering
  \includegraphics[width=8cm]{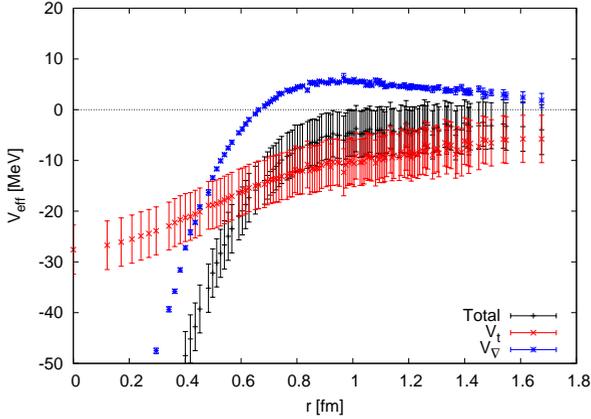}
  \caption{
    \label{fig:pot_each_term}
    Each term of the $J/\psi N$ ($J=1/2$) potential in
    Eq.~\eqref{eq:each_term}: (red)$V_t(r)$, (blue)$V_{\nabla}(r)$,
    and (black)$V_{\text{eff}}(r)$ at $t=12$.
  }
\end{figure}

Figure~\ref{fig:pot_each_term} shows each term of the $J/\psi N$ ($J=1/2$)
potential in Eq.~\eqref{eq:each_term}, depicted separately.
The second term $V_{\nabla}(r)$ gives dominant contribution to the
short-distance attraction, and it behaves repulsively for $r>0.6$.
The first term $V_t(r)$ is attractive overall with smaller strength at
short distances than $V_{\nabla}(r)$.  At large distances, the small
repulsion of $V_{\nabla}(r)$ and the small attraction of $V_t(r)$
cancel and give the totally attractive and short-ranged
$V_{\text{eff}}(r)$.
Notice that $V_t(r)$ is dependent on $r$; it means that the
ground-state saturation of the four-point correlation function is not
achieved.  The time-dependent method gives a desired potential, as
$V_{\text{eff}}(r)$ is naturally connected to outside of its range.

We have performed a fit of the potentials in Figure~\ref{fig:pot_all}
with two Gaussian functions.
Then by solving the S-wave radial Schr\"odinger equation
\begin{align}
\left\{ 
\frac{1}{2\mu r^2}\frac{d}{dr}\left( r^2 \frac{d}{dr} \right) 
+ E
\right\} R(r;E)
=
V(r) R(r;E)
\end{align}
we obtain the scattering phase shift $\delta(E)$ from the asymptotic
form of $R(r;E)$,
\begin{align}
R(r;E) \sim 
\frac{i}{2} \left(
h_0^{(-)} (kr) - e^{2i\delta(E)} h_0^{(+)}(kr),
\right)
\end{align}
where $h_0^{(\pm)}(z)$ are the spherical Riccati-Hankel functions with
$l=0$.

\begin{figure}[thb]
  \centering
  \includegraphics[width=8cm]{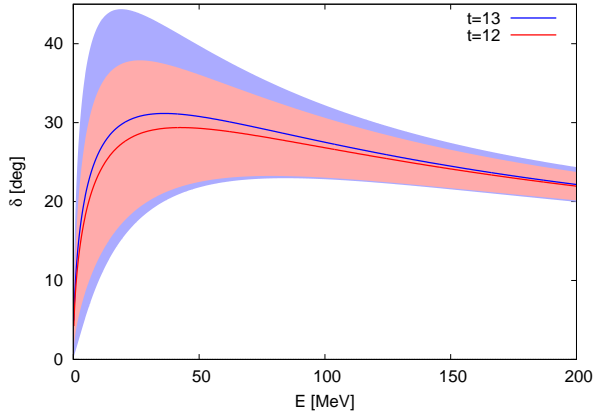}
  \caption{
    \label{fig:phase_shift}
    The scattering phase shift of the $J/\psi N$ ($J=1/2$) at $t=12$ and $13$.
  }
\end{figure}

Figure~\ref{fig:phase_shift} shows the result of the phase shift of 
$J/\psi N$ ($J=1/2$).
We see that the interaction is attractive, but not strong enough to
have bound states.  
Moreover we have performed the effective-range expansion
$k\cot(\delta)=1/a+rk^2/2$ to obtain the scattering length $a$ and the
effective range.  At $t=12$, the results are:
$a=0.68\pm0.44\mathrm{fm}$ and $r=1.04\pm0.03\mathrm{fm}$ for $J/\psi
N$ ($J=1/2$), 
$a=0.63\pm0.42\mathrm{fm}$, $r=1.11\pm0.03\mathrm{fm}$
for $J/\psi N$ ($J=3/2$), 
and $a=0.44\pm0.34\mathrm{fm}$,
$r=1.33\pm0.06\mathrm{fm}$ for $\eta_c N$.

\section{Conclusions}\label{sec:conclusions}

We have computed the S-wave effective central potentials of the
$J/\psi N$ ($J^P=1/2^-$ and $3/2^-$) and $\eta_c N$.
We have shown that our results are qualitatively similar to the
previous results by Kawanai and Sasaki, where the time-independent
method is employed.
However, we have confirmed that the time-dependent method is crucial
to extract a reliable potential, as the ground-state saturation is not
achieved at a moderate time.
Our results show that all the charmonium-nucleon potentials are quite
similar, and this is naturally understood through the heavy quark
symmetry.
We have not observed any bound states as shown in the phase shift
analysis.  However, inclusion of the other channels might result in
emergence of resonance states, which may correspond to the $P_c$
resonances.
The analysis with coupled channels and the use of the RHQ action are
currently in progress.

\vspace{2cm}

\begin{acknowledgement}
{\large{\bf Acknowledgements}}\\ 
This work is supported by Japan Society for the Promotion of Science
KAKENHI Grands No. JP25400244 and by Ministry of Education, Culture,
Sports, Science and Technology as ``Priority Issue on Post-K computer''
(Elucidation of the Fundamental Laws and Evolution of the Universe)
and Joint Institute for Computational Fundamental Science.
We thank CP-PACS and JLQCD Collaborations~\cite{Ishikawa:2007nn} and
ILDG/JLDG~\cite{ILDG/JLDG} for the 2+1 flavor QCD gauge
configurations.  This work is in part based on Bridge++
code~\cite{Bridge}.
\end{acknowledgement}


\end{document}